\documentclass[11pt]{article}
\usepackage[hmargin={2cm,3cm},vmargin={2cm,2cm}]{geometry}
\usepackage{cite}
\usepackage{relsize}
\usepackage{amsmath,amssymb}
\usepackage{mathtools}
\usepackage{hyperref}
\usepackage{amsfonts}
\usepackage{mathrsfs}
\usepackage{enumerate}
\usepackage{color}
\usepackage{marginnote}
\usepackage{epic}
\usepackage{eepic}
\usepackage{graphicx}

\usepackage{tikz} 
\usetikzlibrary{fadings}
\usetikzlibrary{arrows}
\usepackage[normalem]{ulem}
\DeclareMathAlphabet{\mathpzc}{OT1}{pzc}{m}{it}
\DeclareMathAlphabet\mathbfcal{OMS}{cmsy}{b}{n} 
\makeatletter
\@addtoreset{equation}{section}
\makeatother

\newcounter{app}
\newcounter{sapp}[app]

\usepackage{cancel}

\usepackage{graphicx}
\usepackage[small,bf
]{caption}
\renewcommand{\author}[1]{\large\rm #1\\ \bigskip}
\newcommand{\address}[1]{{\normalsize\it #1\\}\bigskip}
\renewcommand{\title}[1]{\bigskip\bigskip\Large\bf #1\bigskip\bigskip\\}

\newcommand{\ds}{\displaystyle}

\def\EXP{\textrm{{\large e}}}

\newcommand{\sx}{{x}}

\newcommand{\xop}{\boldsymbol{x}}

\newcommand{\as}{{\phi}}

\newcommand{\rop}{\mathsf{R}}
\newcommand{\ii}{\mathsf{i}}
\newcommand{\ri}{{\rm i}}
\newcommand{\rd}{{\rm d}}
\newcommand{\re}{\mathsf{e}}
\renewcommand{\re}{\mbox{e}}

\newcommand{\bb}{\mathsf{b}}

\newcommand{\F}{{\mathsf f}}
\newcommand{\G}{{\mathsf G}}

\newcommand{\Fop}{{\mathbf F}}

\newcommand{\Rop}{{\mathbf R}}

\newcommand{\Hi}{{\mathcal H}}
\renewcommand{\S}{{\mathcal S}}
\newcommand{\ph}{\varphi}

\newcommand{\be}{\begin{equation}}
\newcommand{\ee}{\end{equation}}
\newcommand{\bea}{\begin{eqnarray}}
\newcommand{\eea}{\end{eqnarray}}

\newcommand{\sk}[1]{\G (#1)}

\newcommand{\go}{{\gamma}}

\begin{document}

\vglue 2cm

\begin{center}

\title{Quantum Dilogarithms and New Integrable Lattice Models in Three
  Dimensions.}

\vspace{.5cm}

\author{Vladimir V.~Bazhanov$^{1}$, Rinat M.~Kashaev$^{2}$, Vladimir V.~Mangazeev$^{1}$\\ and Sergey M.~Sergeev$^{1,3}$}

\vspace{.5cm}

\address{
$^1$Department of Fundamental \& Theoretical Physics,
Research School of Physics,\\
Australian National University, Canberra, ACT 2601, Australia.\\
\ \\
$^2$ Section de Math\'ematiques
Universit\'e de Gen\`eve,\\ 
1211 Gen\`eve 4, Suisse. \\
\ \\
$^3$Faculty of Science \& Technology,\\
University of Canberra, Bruce, ACT 2617, Australia.}

\abstract
In this paper we introduce a new class of integrable 3D lattice
models, possessing continuous families of commuting layer-to-layer
transfer matrices.  
Algebraically, this commutativity is based on 
a very special construction of local Boltzmann weights in terms of
quantum dilogarithms satisfying the inversion and 
pentagon identities. 
We give three examples of such quantum dilogarithms,  
leading to integrable 3D lattice models.
The partition function per site in these models can be exactly calculated 
in the limit of an infinite lattice by using the functional relations,
symmetry and factorization properties of the transfer matrix. 
The results of such
calculations for 3D models associated with the Faddeev modular
quantum dilogarithm are briefly presented.

\end{center}

\vspace{1cm}
\newpage

\setcounter{tocdepth}{3}
\newpage


\section{Introduction}

The Zamolodchikov tetrahedron equation (ZTE) \cite{
Zamolodchikov:1980rus,Zamolodchikov:1981kf} is the
three-dimensional analogue of the Yang–Baxter equation. It implies the
commutativity of layer-to-layer transfer matrices for
three-dimensional lattice models \cite{Bazhanov:1981zm}
and, thereby, generalizes the most fundamental integrability
structure of exactly solvable lattice models in two dimensions 
\cite{Bax82}. 
The first solution of the tetrahedron equation was presented by
Zamolodchikov \cite{Zamolodchikov:1980rus,Zamolodchikov:1981kf}
 and later proven by Baxter \cite{Baxter:1983qc}. 
Subsequently, many other interesting results in this field have 
been obtained by various authors, see
\cite{
Bazhanov:1981eg,Baxter:1986phd,Maillet:1989gg,
Bazhanov:1992jq,Bazhanov:1993j,
Kashaev:1993ijmp,
Korepanov:1993jsp,Bazhanov:1993pa,Hietarinta:1994pt,  
Kapranov94,
Sergeev:1995rt,
Kashaev:1996on,
Baxter:1997tw,
Sergeev:1998so,
MR1637789,
Kashaev:2000fr,
Bazhanov:2005as,
BMS08a,
Sergeev:2008zu,
Inoue:2023rer}.

Obviously, any integrable
three-dimensional model on a cubic lattice can be viewed as a
two-dimensional model on a square lattice, where the additional third
dimension is treated as an internal degree of freedom.  Therefore
every solution of the tetrahedron equation provides an infinite
sequence of solutions of the Yang-Baxter equation differing by the
size of this ``hidden third dimension''. 
In this way the tetrahedron equation becomes connected to the
Yang-Baxter equation associated with quantized (affine) Lie algebras
(see, e.g., \cite{Bazhanov:1992jq,Kashaev:2000fr,Bazhanov:2005as,KOS:2014} 
for various examples of such connections). As a result, 
the tetrahedron equation becomes connected to the theory of Quantum
Groups, but, nevertheless, 
its underlying algebraic structure still appears to be
not entirely understood.
Therefore, it is important to explore alternative
approaches to integrability in three dimensions, which are based on
other algebraic relations, simpler than the tetrahedron equation.

In this paper we continue to develop one such scheme, originally
introduced in \cite{Baxter:1986phd} 
and further extended in \cite{Bazhanov:1993j,Sergeev:1995rt,Sergeev:2004pn}.  
Consequently, we 
present 
a new class of integrable three-dimensional lattice models. 
These models admit various equivalent formulations, in particular, as  
``vertex models'' with fluctuating ``spin variables'' (or just
``spins'') placed at the edges of the cubic lattice and local
Boltzmann weights assigned to the lattice vertices. Alternatively, the same
models can be reformulated as ``interaction-round-a-cube'' (IRC) models 
with spins placed at the lattice sites and local Boltzmann weights
assigned to the corner 
spin configurations surrounding elementary lattice cubes. 
In both cases the associated Boltzmann weights have very special form, 
so that the models can be viewed as Ising-type models 
with two- and three-spin interactions only.
These elementary interactions  
obey a number of special
relations, which ensure the commutativity of the layer-to-layer
transfer matrices and the exact integrability of the model.  The most
important among these relations is the ``restricted star-triangle
relation'' (RSTR) of Bazhanov and Baxter \cite{Bazhanov:1993j}, 
which relates the two-
and three-spin weights.  Remarkably, the very same relation\footnote{%
To be more precise, the restricted star-triangle
relation of \cite{Bazhanov:1993j}
 is slightly more general than the pentagon identity 
of \cite{Faddeev:1993rs},
and sometimes referred to as the ``parameter-dependent pentagon
identity'' (see \cite{bkms2} more details).}
serves as the five-term (or the ``pentagon'') identity for the quantum
dilogarithm introduced by Faddeev and Kashaev
\cite{Faddeev:1993rs}.
Therefore, each example of
the quantum dilogarithm leads to a new 3D integrable lattice
model. This is the main connection, which we will exploit and develop
in the present paper.

The five-term identity implies a number of 
subsidiary
relations, including the ``3D star-star'' relation
\cite{Bazhanov:1992jq, Bazhanov:1993j, Kashaev:1993ijmp},  
and the tetrahedron equation, which both lead to the commutativity of transfer 
matrices. Moreover, due to the special structure of the Boltzmannn
weights, the partition function per site for, at least, some of these models 
can be exactly calculated in the limit of an infinite lattice by using the
symmetry and factorization properties, 
as it was originally done by Baxter \cite{Baxter:1986phd} 
for the Zamolodchikov model. 
Therefore, the five-term identity (or 
the restricted star-triangle
relation) can be regarded as a primary integrability condition for
this class of 3D lattice models. 

In this paper we consider three-dimensional models associated with
different examples of the quantum dilogarithm. The lattice spin
variables in each of these models run over some specific (model
dependent) set of values $\S$. In particular, we present new 
models, where the spins take their values  
\begin{enumerate}[(a)]
\item
on the real line $\S={\mathbb R}$ (associated with the Faddeev
dilogarithm \cite{Faddeev:1994fw}),  
\item
on the direct product $\S={\mathbb R}\times {\mathbb Z}_N$, where
${\mathbb Z}_N$ is the set of integers \emph{modulo} $N$ (associated with
Andersen--Kashaev dilogarithm \cite{Andersen:2014aoa}), 
\item
on the direct product $\S={\mathbb T}\times {\mathbb Z}$, where
${\mathbb T}$ is the unit circle and ${\mathbb Z}$ is the set of all
integers (associated with Woronowicz dilogarithm 
\cite{Woronowicz:1992}).  
\end{enumerate}
More generally, we give universal formulae expressing solutions of 
the Zamolodchikov tetrahedron equation in terms of quantum dilogarithms 
(both for the ``interaction-round-a-cube'' and the vertex-type forms 
of the equation). 

The organisation of the paper is as follows. In Sect.\ref{vertex-sect} 
we introduce the quantum dilogarithms and describe their main properties
requred for construction of integrable 3D lattice models. For each
example of the quantum dilogarithm we present three different, but
related, solutions of the Zamolodchikov tetrahedron equation (of the
vertex and interaction-round-a-cube forms). 
Next in Sect.\ref{examples} we review the three examples of the quantum
dilogarithm, mentioned above.
Finally, in Sect.\ref{partition} we
present results of exact calculation of partition functions for 
3D lattice models, associated with the Faddeev dilogarithm. The
details of all derivations will be published separately \cite{bkms2}.

\section{Solvable lattice models and quantum dilogarithms
\label{vertex-sect}}
In this paper we consider a particular class of 
integrable models on three-dimensional 
cubic lattices. These models admit several equivalent
formulations, revealing different aspects of their algebraic structure. 
We start with the {\em vertex models},\/ where the local Boltzmann
weights are associated with 
edge spin configurations around the vertices of the lattice. 
Next, we will also present an equivalent 
{\em interaction-round-a-cube} (IRC) formulation of the same 3D models.

\subsection{Vertex interaction models}
Consider a simple cubic lattice $\cal L$ of the size 
$M\times M\times M$\/ with the periodic
boundary conditions in each direction.  At each edge of $\cal L$ place a
spin variable $\sx$, taking some set of values 
$\sx\in\mathcal{S}$, 
and allow all possible interactions of the spins 
at each vertex of the lattice. 
The partition function reads
\be\label{Zvertex}
Z^{\rm(vert)}=\sum_{\rm spins}\ \  \prod_{\rm vertices} \ \langle
\sx_1,\sx_2,\sx_3|\,\Rop\,|\sx_1',\sx_2',\sx_3'\rangle\,,
\ee  
where $\langle\sx_1,\sx_2,\sx_3|\,\Rop\,|\sx_1',\sx_2',\sx_3'\rangle$ 
denotes the
Boltzmann weight corresponding to the configuration of six edge spins,
arranged as in \eqref{cross} below 
\begin{equation}\label{cross}
\begin{tikzpicture}[scale=2,baseline=(current  bounding  box.center)]
\begin{scope}[shift={(-.4,0)}]
\draw [-latex, ultra thick] (0,-0.75) -- (0,0.75); \node [above] at (0,0.75) {$\sx_1'$}; \node [below] at (0,-0.75) {$\sx_1$};
\draw [-latex, ultra thick] (-1,0) -- (1,0); \node [right] at (1,0) {$\sx_2'$}; \node [left] at (-1,0) {$\sx_2$};
\draw [-latex, ultra thick] (0.8,0.3) -- (-0.8,-0.3); \node [right] at (0.8,0.4) {$\sx_3$}; \node [left] at (-0.8,-0.4) {$\sx_3'$};
\end{scope}
\node [left] at (-1.8,0) 
{$\langle
  \sx_1^{},\sx_2^{},\sx_3^{}|\,\Rop\,|\sx_1',\sx_2',\sx_3'\rangle\;=$};  
\end{tikzpicture}
\end{equation}
Evidently, we cannot solve the model \eqref{Zvertex} for an arbitrary
Boltzmann weight function. 
But what we, apparently, can do is to solve
a restricted class of models whose layer-to-layer 
transfer matrices form continuous commuting families. 
In all known cases such commutativity condition can always be
reformulated as a relation for the local Boltzmann weights. 
In two dimensions the corresponding relation is
well known: it is the Yang-Baxter equation \cite{Bax82}.
A generalization of this relation 
to a 3D lattice is also known \cite{Bazhanov:1981zm}: it is the Zamolodchikov 
tetrahedron equation
\cite{Zamolodchikov:1980rus,Zamolodchikov:1981kf}.  
Here we construct new solutions of 
this complicated equation by reducing it to much simpler algebraic
relations, satisfied by the quantum dilogarithm.   
To describe these results we need to introduce new notions and notations.

\bigskip
\subsection{Quantum dilogarithm\label{s22}}
Below we introduce 3D lattice models where the spins 
take infinitely many
values. In particular, we will consider three different cases, where the edge
spins $\sx \in {\S}$ take their values in 
\be\label{P-list}
\S={\mathbb R}\,,\qquad {\S}={\mathbb T}\times {\mathbb Z}\,,\qquad 
{\S}={\mathbb R}\times {\mathbb Z}_N\,.
\ee
Here 
${\mathbb R}$ denotes the real line, ${\mathbb T}$ the unit circle,
${\mathbb Z}$ the set of all integers and ${\mathbb Z}_N$  ($N\ge2$) denotes 
the set of integers {\em modulo} $N$. It is useful to note, that the
above sets \eqref{P-list} are examples of the {\em Pontryagin
  self-dual locally compact Abelian (LCA) groups}, which we simply
call as Pontryagin groups (see \cite{MR0005741} for additional details). 

Let ${\mathcal
  H}=L^2(\S)$ denote the Hilbert space of quadratically integrable
(complex valued) functions on $\S$,
\be
\Psi\in  L^2(\S):\qquad\qquad \int_\S\,|\Psi(x)|^2\,\rd x<\infty\,,\qquad
\qquad x\in \S\,,
\ee
where the integral sign $\int_{\S}$ stands for a (normalized) 
sum over discrete components of the variable $x\in {\S}$ 
and an integral over its continuous components (detailed
conventions for each case of ${\S}$ are given in Sect.\ref{examples}).
It will be convenient to use the Dirac bra-ket notations for 
the elements of $\S$,  
\be\label{coor-gen}
\langle x|\Psi(\xop) =\langle x|\Psi(x)\,,\qquad
\langle x|x' \rangle=\delta(x-x')\,,\qquad 1=\int_{\S}| x\rangle
\,\rd x\,\langle x|\,, \qquad x,x'\in\S\,.
\ee
where we have introduced the 
coordinate operator $\xop$ acting diagonally on
the basis vectors. As usual, $\delta(x)$ denotes the delta-function.
Note, that since every considered set $\S$ is an Abelian group with respect to
addition, the sum of two elements $x,x'\in\S$ (or their difference)
also belongs to the same set $x\pm x'\in\S$.

Next, we need to define the Fourier transformation, which maps the
Hilbert space
$L^2(\S)\to L^2(\S)$ to itself. There is a unitary
Fourier transform operator ${\mathbf F}$ 
which can be defined for each $\S$ in \eqref{P-list} 
\be\label{F-def}
\langle x |\Psi\rangle={\Psi}(x)\,,\qquad
\langle x |{\mathbf F}\Psi\rangle=\widetilde{\Psi}(x)=
\int_{\S}\,\F(x,y)\,\Psi(y)\,\rd y\,,\qquad
x\in{\S},\qquad \forall\, \Psi\in L^2({\S})\,,
\ee
such, that the integral kernel of this transform 
\be\label{kernel}
\langle x|{\mathbf F}| y \rangle=\F(x,y)=\F(y,x)
\,,\qquad \F(x,y+z)=\F(x,y)\, \F(x,z)\,,
\ee
satisfies the conditions
\be\label{delta}
\int_{\S}\,\F(x,y)\,\rd y=\delta(x)\,,\qquad 
\int_{{\S}^2}\,\F(x,y)\,\rd x\,\rd y=1\,.
\ee
%
The last step is to introduce the Gaussian exponent $\G(x)$,
satisfying the functional relations, 
\be\label{gauss-def}
\G(x)=\G(-x)\,,\qquad \G(x+y)=\F(x,y)\,\G(x)\,\G(y)\,, 
\ee
with the Fourier kernel $\F(x,y)$. By elementary manipulations with 
\eqref{kernel} and \eqref{gauss-def} it is easy to deduce, that
\be
\F(x,0)=\F(0,x)=\G(0)=1\,,\qquad \F(x,x)=\G(x)^2\,,\qquad 
x\in{\mathcal S}\,.
\ee

\bigskip
We can now define a quantum dilogarithm over 
$\S$ as a
complex valued function $\varphi(x)$, $x\in\S$,
which satisfies the following relations \cite{Andersen:2014aoa,Kashaev:2015}   
\begin{enumerate}[(i)]
\item
\uline{{\sl inversion relation}}  
\be\label{inv-gen}
\varphi(x)\varphi(-x)\;=\;\varphi(0)^2\, \G(x)\,,\qquad \varphi(0)\not=0\,,
\qquad \forall\;x\in{\mathcal S}\;,
\ee
\item
\uline{{\sl the quantum five-term identity}}
\be\label{five1}
\varphi(z-x)\,\widetilde{\varphi}(x-y)\,\varphi(y-z)=
\varphi(0)^2\,\int_{\S}
\widetilde{\varphi}(x-w)\,\G(w-z)\,\,\widetilde{\varphi}(w-y)\,\rd w 
\ee
where the tilde mark in $\widetilde{\varphi}$ stands for the 
Fourier transformation,
\be\label{Fourier-def0}
\widetilde{\varphi}(x)=\int_{\mathcal S}\,\F(x,y)\,\varphi(y)\,\rd y\,.
\ee
Remembering the definition of the coordinate operator $\xop$
in \eqref{coor-gen} and 
using \eqref{kernel} and \eqref{gauss-def}, it is easy to
rewrite the five-term identity \eqref{five1} in the operator form
\be\label{pentagon-gen}
\big({\mathbf F}\,\varphi(\xop)\,{\mathbf F}^{-1}\big)\,\,\varphi(\xop)\,=
\varphi(\xop)\,\,\big({\G(\xop)}^{-1}\,{\mathbf F}\,\varphi(\xop)\,
{\mathbf F}^{-1}\,{\G(\xop)}\big)\,\,\big({\mathbf F}\,\varphi(\xop)\,{\mathbf F}^{-1}\big)\,,
\ee
Indeed, calculating the matrix
elements of both sides of the last relation between the states $\langle
x|$ and $|y\rangle$ and using \eqref{inv-gen}, one immediately obtains
\eqref{five1} with $z=0$.
\end{enumerate}

\bigskip
The above relations \eqref{inv-gen} and \eqref{five1} are the main 
properties defining the notion of the quantum dilogarithm. 
For some
applications (e.g., for constructing the solution 
\eqref{elem-gen} of the tetrahedron
equation \eqref{ZTE-vertex}) 
no further properties are required.
However, 
below 
we will use 
quantum dilogarithms possessing an additional property, which will be 
called as  
\begin{enumerate}[(iii)]
\item
\uline{{\sl Fourier transform self-duality}}  
\be\label{four-gen}
\widetilde{\varphi}(x)=\int_{\mathcal S}\,\F(x,y)\,\varphi(y)\,\rd y=
\go\,\varphi(0)^2\,\G(x)^{-1}\,\varphi(x+\eta)\,,
\qquad \go=\int_{\mathcal S} \G(x)\,\rd x\,,
\ee
where $\varphi(0)$, $\eta$ and $\gamma$ are $\S$-dependent  
constants, which are constrained as\footnote{%
Eq. \eqref{parrel} is a necessary constraint arising as a consistency condition of the Fourier transformation formula \eqref{four-gen} with the inversion relation \eqref{inv-gen}.} 
\be\label{parrel}
\go^2\,\varphi(0)^6\,\G(\eta)=1\,.
\ee
\end{enumerate}
Using the last relation and \eqref{inv-gen} one can write
\eqref{four-gen} in an equivalent form
\begin{equation}\label{four-gen2}
\frac{1}{\varphi(x-\eta)}=
{\gamma\,\varphi(0)^2}\,{\G(x)^{-1}}\,
\int_\S \,\frac{dy}{\F(x,y)\,\varphi(y)}
\end{equation}
Note, that in general the constant 
$\eta$ does not belong to the set $\S$. 
Therefore, the ``shifted functions''
 $\varphi(x\pm\eta)$ in \eqref{four-gen} and \eqref{four-gen2} 
should be understood as an analytic continuation of $\varphi(x)$ from
$x\in \S$. Actually, any of these equations could be thought as
a definition of the constant $\eta$, which we will sometimes call 
as the ``crossing parameter''.  

Using \eqref{inv-gen} and 
\eqref{four-gen} in the five-term 
identity \eqref{five1}
and then replacing  
$x\to x-y-\eta$, $y\to \eta$, $z\to -z$  and $w\to w+x$, one obtains
another convenient form of \eqref{five1}, 
\begin{subequations}\label{fterm3}
\begin{equation}\label{fterm3a}
\int_{\mathcal S} \frac{{\varphi}(w+x)}{{\varphi}(w+y)}
\,\F(z,w)\,\rd w\;=\;
\frac{\go\,{\varphi}(0)^2}{\F(z,y+\eta)}
\frac{{\varphi}(x-y-\eta)\,
{\varphi}(z+\eta)}{{\varphi}(x-y+z-\eta)}\;.
\end{equation}
or, equivalently,
\begin{equation}\label{fterm3b}
\int_{\S} \frac{\ph(w+x)}{\ph(w+y)}\F(z,w)\rd w\;=\;
\frac{\F(z,\eta-x)}{\go\varphi(0)^2} 
\frac{\ph(-x+y-z+\eta)}{\ph(-x+y+\eta)\ph(-z-\eta)}\;.
\end{equation}
\end{subequations}

\bigskip\noindent

It should be stressed, that, 
even though we will be considering the spin sets  
\eqref{P-list} as our main examples, 
all the above reasonings leading to
\eqref{pentagon-gen}, 
\eqref{four-gen2} and \eqref{fterm3} 
are rather general and equally apply to any set $\S$, 
as long as it allows to 
define the quantum dilogarithm possessing the properties
\eqref{inv-gen}-\eqref{parrel}.

The reader might have noted, that the relation \eqref{five1} strongly
resembles the Onsager's star-triangle relation (see, e.g.,
\cite{Bax82}). Actually, it is a particular case of the so-called
restricted star-triangle relation \cite{Bazhanov:1993j}, which has
appeared as an integrability condition for the $N$-state
generalization \cite{Bazhanov:1992jq} of the 3D Zamolodchikov model
\cite{Zamolodchikov:1980rus,Zamolodchikov:1981kf}. We will further discuss this connection in our next paper \cite{bkms2}.  

\subsection{Vertex-type solutions of the Zamolodchikov tetrahedron equations}
With the above definitions of the quantum dilogarithms we could now
finalize the formulation of our 3D vertex model
\eqref{Zvertex}. Consider the following general expression for
the Boltzmann weights \eqref{cross}
\be\label{elem-gen}
\begin{array}{l}
\ds \langle x_1,x_2,x_3|\, \Rop\, | x_1',x_2',x_3'\rangle 
\;=\;
\ds \rho\;\delta(x_2+x_3-x_2'-x_3')\,
\frac{\sk{\lambda_1}\,\sk{x_2}}{\sk{\lambda_3}\,\sk{x_2'}} 
\frac{\F(x_1,x_3)}{\F(x_1',x_3')}\,\times \\
\\
\times\;
\ds\F(x_1'-x_2-\lambda_2,\lambda_1-\lambda_3)\,
\frac{\ph(\lambda_2-x_1'+x_2')}{\ph(\lambda_1+x_1-x_2)}\,
\widetilde{\ph}(\lambda_1-\lambda_2+x_1'-x_2) \,
\overline{\widetilde{\ph}}(x_2'-x_1)\,,\\
\end{array}
\ee
where 
\be\label{tildas}
\widetilde{\ph}(x)\;=\;\int \F(x,y)\,\ph(y) \,dy\;,\qquad
\overline{\widetilde{\ph}}(x)\;=\;\int \frac{dy}{\F(x,y)\, \ph(y)}\;.
\ee
where $\lambda_1,\lambda_2,\lambda_3\in\S$ are arbitrary
parameters and 
$\rho=\rho(\lambda_1,\lambda_2,\lambda_3)$
is a
scalar normalization factor, introduced for  
the later convenience.  All other notations, including those for 
the Fourier kernel $\F(x,y)$, the Gaussian
exponent $\G(x)$ and the quantum dilogarithm $\varphi(x)$ were defined
in the previous subsection. Note, that the definition 
\eqref{elem-gen} only requires  
the basic properties $(i)$ and  $(ii)$ of the dilogarithm 
$\varphi(x)$, given by \eqref{inv-gen} and \eqref{five1}.

Following common tradition we will call the expression
\eqref{elem-gen} as an $\Rop$-matrix. 
This quantity could conveniently be
associated with a linear operator 
\be\label{H-cube}
\Rop_{123}(\lambda_1,\lambda_2,\lambda_3):\qquad \Hi\otimes\Hi\otimes\Hi\to
\Hi\otimes\Hi\otimes\Hi\,,
\ee
acting in the tensor product of three Hilbert spaces $\Hi=L^2(\S)$,
 such that the pairs
of indices $(\sx_1,\sx_1')$, $(\sx_2,\sx_2')$ and $(\sx_3,\sx_3')$ serve as
matrix indices in the first, second and third spaces,
respectively. Similarly, the parameters 
$\lambda_1,\lambda_2,\lambda_3\in\S$, entering \eqref{elem-gen},
are also associated, respectively, with the
first, second and third Hilbert spaces in \eqref{H-cube}. 

We claim, that the $\Rop$-matrix \eqref{elem-gen} satisfies the
vertex-type Zamolodchikov tetrahedron equation 
\bea\nonumber
\Rop_{123}(\lambda_1,\lambda_2,\lambda_3)\,
\Rop_{145}(\lambda_1,\lambda_4,\lambda_5)\,
\Rop_{246}(\lambda_2,\lambda_4,\lambda_6)\,
\Rop_{356}(\lambda_3,\lambda_5,\lambda_6)=\qquad\qquad&&\\[.4cm]
=\Rop_{356}(\lambda_3,\lambda_5,\lambda_6)\,
\Rop_{246}(\lambda_2,\lambda_4,\lambda_6)\,
\Rop_{145}(\lambda_1,\lambda_4,\lambda_5)\,
\Rop_{123}(\lambda_1,\lambda_2,\lambda_3)\,,&&\label{ZTE-vertex}
\eea
containing six independent spectral parameters $\lambda_1,\ldots,\lambda_6$.
The above equation involves operators acting
in the tensor product ${\mathcal H}^{\otimes 6}$ of six identical 
infinite-dimensional Hilbert spaces ${\mathcal H}=L^2(\S)$, where 
the $i$-th space is associated with the corresponding parameter $\lambda_i$.
The operator ${\Rop}_{ijk}(\lambda_i,\lambda_j,\lambda_k)$ acts nontrivially  
in the $i$-th, $j$-th and 
$k$-th spaces, and acts as the identity operator in the other three spaces.
As is well known \cite{Bazhanov:1981zm} the equation \eqref{ZTE-vertex} 
implies the commutativity of layer-to-layer transfer matrices for the
vertex model \eqref{Zvertex}.

The proof of the tetrahedron equation \eqref{ZTE-vertex} 
is given  in \cite{bkms2}.
It does not require a
specialization to a particular set of spin variables in
\eqref{P-list}. It is only based on the basic properties of the
quantum dilogarithms, given by \eqref{inv-gen} and \eqref{five1}.

Note, that for the Fourier self-dual dilogarithms,
possessing the additional property \eqref{four-gen},  the expression 
\eqref{elem-gen} could be simplified as 
\begin{equation}\label{R-matrix2}
\begin{array}{l}
\langle\sx_1,\sx_2,\sx_3|\,\Rop\,|\sx_1',\sx_2',\sx_3'\rangle
\,=\ds \rho_1\,\delta(x_2+x_3-x_2'-x_3')\,
\F(x_3'-\lambda_1,x_1-x_1')\, \times\\
[6mm]
\hspace{-0em}\ds\qquad\quad\times \,\F(\lambda_3-\eta, x_2-x_1')\,
 \frac{ \ph(-\lambda_1-x_1+x_2)\,
 \ph(\lambda_2-x_1'+x_2')_{\phantom{|_1}} }
{ \ph(-\eta-x_1+x_2')\,
 \ph(-\eta-\lambda_1^{}+\lambda_2-x_1'+x_2) }
\end{array}
\end{equation}
where the parameter $\eta$ is defined by \eqref{four-gen} and 
$\rho_1$ is a scalar factor 
\be\label{varrho-def}
\rho_1(\lambda_1,\lambda_2,\lambda_3)\;=\;
\rho(\lambda_1,\lambda_2,\lambda_3)\,\G(\eta)\, 
\F(\eta,\lambda_1-\lambda_2)\, \F(\lambda_2,\lambda_3-\lambda_1)/
\G(\lambda_3)\;,
\ee
which is proportional to the 
factor $\rho$, introduced in \eqref{elem-gen}. Looking ahead to the
consideration of the symmetry properties of the model in our next
paper \cite{bkms2},
it is convenient to set the following normalization
of the vertex weights \eqref{R-matrix2} 
\be\label{ro-1}
\rho_1(\lambda_1,\lambda_2,\lambda_3)=
\big(\F(\eta-\lambda_3,\eta+\lambda_1-\lambda_2)
/\F(\lambda_1,\lambda_2)\big)^{\frac{1}{2}}\,.
\ee

Note, that some particular cases of the solution (2.23) were
previously obtained   
in \cite{Sergeev:2009jgp} (for the Faddeev dilogarithm), as well as in  
\cite{Sergeev:1995rt} and \cite{Sergeev:1999jpa} for the cyclic and 
compact quantum dilograrithms (the latter two are not considered here).

\subsection{Interaction-round-a-cube models\label{IRC}}
The vertex model, 
defined by \eqref{Zvertex} and \eqref{R-matrix2},
could be equivalently reformulated as an ``interaction-round-a-cube'' (IRC)
model with spins assigned to the sites of the dual lattice  
and the local Boltzmann weights attributed to configurations
of spins $a,b,c,d,e,f,g,h\in\S$
around each elementary cube, arranged as shown below.%
\footnote{Note, that our arrangement of the corner spins $a,b,c,d,e,f,g,h$
is slightly different from that used in
\cite{Baxter:1986phd, Bazhanov:1992jq, Bazhanov:1993j}. 
The two arrangements are obtained from each other by a   
  reflection about the plane passing through the 
corners  $a,e,h,b$. Fortunately, this only affects the visual
  appearance, but does not influence the associated formulae.}
\be\label{IRC-cube}
\begin{tikzpicture}[scale=1.3,baseline=(current  bounding  box.center)]
\def\bcoler{blue!100}
\def\acoler{black!60}
\draw [\bcoler,fill] (1.8,1+0.3) circle [radius=0.04];
\draw [\bcoler,fill] (0.2,1-0.3) circle [radius=0.04];
\draw [\bcoler,fill] (-1.8,1-0.3) circle [radius=0.04];
\draw [\bcoler,fill] (-0.2,1+0.3) circle [radius=0.04];
\draw [\bcoler,fill] (1.8,-1+0.3) circle [radius=0.04];
\draw [\bcoler,fill] (0.2,-1-0.3) circle [radius=0.04];
\draw [\bcoler,fill] (-1.8,-1-0.3) circle [radius=0.04];
\draw [\bcoler,fill] (-0.2,-1+0.3) circle [radius=0.04];
\draw [\bcoler, very thick] (1.8,1+0.3) -- (0.2,1-0.3) -- (-1.8,1-0.3) -- (-0.2,1+0.3) -- (1.8,1+0.3);
\draw [\bcoler,very thick] (1.8,1+0.3) -- (1.8,-1+0.3) -- (0.2,-1-0.3) -- (0.2,1-0.3);
\draw [\bcoler,very thick] (0.2,-1-0.3) -- (-1.8,-1-0.3) -- (-1.8,1-0.3);
\draw [\bcoler,dashed] (-0.2,1+0.3) -- (-0.2,-1+0.3) -- (1.8,-1+0.3);
\draw [\bcoler,dashed] (-0.2,-1+0.3) -- (-1.8,-1-0.3);
\node [\bcoler,above] at (1.8,1+0.3) {$g$};
\node [\bcoler,above] at (0.2,1-0.3) {$a$};
\node [\bcoler,above] at (-1.8,1-0.3) {$f$};
\node [\bcoler,above] at (-0.2,1+0.3) {$b$};
\node [\bcoler,below] at (1.8,-1+0.3) {$c$};
\node [\bcoler,below] at (0.2,-1-0.3) {$e$};
\node [\bcoler,below] at (-1.8,-1-0.3) {$d$};
\node [\bcoler,below] at (-0.2,-1+0.3) {$h$};
\node [left] at (-2,0) {$W(a|e,f,g|b,c,d|h)\;=$};
\end{tikzpicture}
\ee
The partition function is defined
as
\be
Z^{\rm{(IRC)}} = \sum_{\rm{spins}} \ \prod_{\rm{cubes}}\,
\,W(a|e,f,g|b,c,d|h).
\label{Z-cube}
\ee
where the Boltzmann weights
\begin{equation}\label{W-im2}
\begin{array}{l}
\ds W(a|e,f,g|b,c,d|h;\,\lambda_1,\lambda_2,\lambda_3) \;=\;\rho_W 
\F(\eta+\lambda_1,c-h)\;
\F(\lambda_1-\lambda_2,b-g)\times\\
\\
\ds \qquad\qquad\qquad\qquad\times
\ {}_2F_2\left(\begin{array}{ll} d-e & b-g-\lambda_1+\lambda_3 \\ h-c-\lambda_1 & f-a+\lambda_3 \end{array} \vline \; \; e+g-a-c-\lambda_2\right)\;,
\end{array}
\end{equation}
are expressed through 
an appropriate analog of the bilateral hypergeometric
series
\begin{equation}\label{Hyper}
{}_2F_2\left(\begin{array}{ll} a_1 & a_2 \\ b_1 & b_2 \end{array} \vline \; \; c\right) \;=\;
\int_{\S} dx \,\F(x,c-\eta)\, \frac{\ph(x+a_1)\,\ph(x+a_2)}{\ph(x+b_1-\eta)\, \ph(x+b_2-\eta)}\;.
\end{equation}
The weights $W$ depend on the eight corner spins and also on the parameters
$\lambda_1,\lambda_2,\lambda_3\in\S$. The scalar factor
$\rho_W$ is defined as
\be\label{ro-w}
\rho_W(\lambda_1,\lambda_2,\lambda_3)=\big(\F(\lambda_1,\lambda_3)\F(\eta+\lambda_2,\eta+\lambda_1-\lambda_3)\big)^{\frac{1}{2}}
\ee
The connection of \eqref{W-im2} with \eqref{R-matrix2} is based on a 3D
generalization \cite{bkms2} of Baxter's ``propagation through the vertex''
technique \cite{Bax73b}, which allows one to relate 
the partition functions \eqref{Zvertex} and \eqref{Z-cube}.
In particular, if the normalization of the IRC weights is set by
\eqref{W-im2} and \eqref{ro-w}, while the normalization of the vertex
weights is set by \eqref{R-matrix2} and \eqref{ro-1}, then for the
periodic boundary conditions both partition functions exactly
coincide. The details of calculations are given in \cite{bkms2}.

Another important consequence of the above 
connection between the vertex and IRC models is that the
weights $W$ satisfy the tetrahedron
equation in the ``interaction-round-a-cube'' form 
\cite{Bazhanov:1981zm,Baxter:1983qc},
\be
\begin{array}{l}
\ds \int_{\S} dz
\, 
W(z|a_3,a_2,a_1|c_5,c_6,c_4|b_4)\,\,
W'(a_4|z,c_3,c_2|b_3,a_1,a_2|c_5)\,
\phantom{X}\\[.3cm]
\qquad\times\, 
W''(c_1|a_3,a_4,b_2|c_2,c_6,z|a_1)\,\,
W'''(b_1|c_4,c_3,c_1,|a_4,a_3,a_2|z)\,
=\phantom{X}\\[.4cm]
\phantom{XXXXX} \ds= 
\int_{{\S}} dz
\, W'''(z|b_4,b_3,b_2|c_2,c_6,c_5|a_1)
\,\, W''(b_1|c_4,c_3,z|b_3,b_4,a_2|c_5)
\\[.3cm]
\qquad\qquad\qquad\qquad\times
\, W'(c_1|a_3,b_1,b_2|z,c_6,c_4|b_4)
\,\, W(a_4|c_1,c_3,c_2|b_3,b_2,b_1|z)\,,
\label{tetr}
\end{array}
\ee 
where, apart from the explicitly indicated spin arguments,
the weights $W, W', W'', W'''$ depend on different sets of spectral parameters
\be
\begin{array}{rcl}
W(\ldots)=W(\ldots;\,\lambda_1,\lambda_2,\lambda_3)\,,\qquad &&
W'(\ldots)=W(\ldots;\lambda_1,\lambda_4,\lambda_5)\,,\\[.3cm]
W''(\ldots)=W(\ldots;\,\lambda_2,\lambda_4,\lambda_6)\,,\qquad &&
W'''(\ldots)=W(\ldots;\lambda_3,\lambda_5,\lambda_6)\,,
\end{array}
\ee
The equation \eqref{tetr} holds for arbitrary values of all six
parameters $\lambda_1,\ldots,\lambda_6\in\S$. The proof is based on
transformation identities for the generalised hypergeometric series.

It is worth noting, that there are simple
equivalence transformations of the weights $W$ 
which do not change the partition function 
\eqref{Z-cube}.  
For instance, if we multiply $W(a,\ldots ,h)$ by
$z(a,g,b,f) /
z(e,c,h,d)$ then each horizontal face of
the lattice acquires a $z$-factor from the cube below it, and a
canceling $(1/z)$ factor from the cube above. 
This is an example of a
``face-factor'' transformation.  Similarly, one could apply
``edge-factor'' and ``site-factor'' transformations that leave
the partition function \eqref{Z-cube} unchanged.

\subsection{A vertex-type $\Rop$-matrix for the IRC model}
\newcommand\mystrut{\rule[-3pt]{0pt}{14pt}}
Interestingly, the $W$ weights \eqref{W-im2}
could be simply converted into a new $\Rop$-matrix (different from
that given by \eqref{R-matrix2}), which satisfies
the vertex-type tetrahedron equation. 
This means, that the IRC model \eqref{Z-cube} can be 
reformulated as yet another vertex model   
without using the ``propagation through the vertex'' transformation. 
Such a connection was explained in \cite{BMS08a}.
The new vertex $\Rop$-matrix reads
\begin{equation}\label{qR1}
\begin{array}{l}
\ds \rop\mystrut_{\sigma_1,\sigma_2,\sigma_3}^{\sigma_1',\sigma_2',\sigma_3'} \;=\; 
\delta(\sigma_1+\sigma_2-\sigma_1'-\sigma_2')\,
\delta(\sigma_2+\sigma_3-\sigma_2'-\sigma_3')\,
\sqrt{\frac{\ph(\sigma_1')\ph(\sigma_2')\ph(\sigma_3')}{\ph(\sigma_1)\ph(\sigma_2)\ph(\sigma_3)}}
\;\times\\
\\
\ds\qquad\qquad \times\;
\frac{\F(\sigma_1,\sigma_3)^{^1\!/\!_2}}{\F(\eta,\sigma_2-\sigma_2')^{^1\!/\!_4}}\;  
{}_2F_2\left(\begin{array}{ll} (\eta+\sigma_1-\sigma_3)/2 &
  (\eta-\sigma_1+\sigma_3)/2 \\ 
(\eta-\sigma_1-\sigma_3)/2 & (\eta+\sigma_1'+\sigma_3')/2 \end{array} \vline -
\sigma_2'\right)
\end{array}
\end{equation}
where 
$\sigma_1,\sigma_2,\sigma_3,\sigma_1',\sigma_2',\sigma_3'\in\S$ 
denote the edge spins. 
To relate the new $\Rop$-matrix with the $W$ weights, 
let us make the substitution 
\begin{equation}\label{sigma}
\begin{array}{ll}
\ds \sigma_1\;=\;c+d-e-h+\lambda_1\;, & \ds
\sigma_1'\;=\;f+g-a-b+\lambda_1\;,\\[.2cm] 
\ds \sigma_2\;=\;f+h-b-d+\lambda_2\;, & \ds
\sigma_2'\;=\;a+c-e-g+\lambda_2\;,\\[.2cm] 
\ds \sigma_3\;=\;b+c-g-h+\lambda_3\;, & \ds \sigma_3'\;=\;e+f-a-d+\lambda_3\;,
\end{array}
\end{equation}
expressing the edge spins via the corner spins $a,b,c,d,e,f,g,h$
arranged as in \eqref{IRC-cube} 
(note, that with this substitution the $\delta$-function constraints in
\eqref{qR1} are automatically satisfied).   
Then, it is not difficult to show that the $\Rop$-matrix \eqref{qR1} 
(with omitted $\delta$-functions) can be written as 
\begin{equation}\label{R-WS}
\rop\mystrut_{\sigma_1,\sigma_2,\sigma_3}^{\sigma_1',\sigma_2',\sigma_3'}\;=\;
\Phi_0\,W(a|e,f,g|b,c,d|h;\lambda_1,\lambda_2,\lambda_3)\;.
\end{equation}
where the weights $W(\ldots)$ are given by \eqref{W-im2} and 
$\Phi_0=\Phi_0(a|e,f,g|b,c,d|h;\lambda_1,\lambda_2,\lambda_3)$ 
is an equivalence transformation factor which cancels out from 
the partition function $\eqref{Z-cube}$ (see the end of Sect.\ref{IRC} for
an example of such factor).   

Note, that the $\Rop$-matrix \eqref{qR1} does not have any spectral
parameters, however, it could be generalized by
introducing the ``external fields'', $\as_1,\as_2,\as_3\in\S$,
\be
\begin{array}{l}
\ds
\rop\mystrut_{\sigma_1,\sigma_2,\sigma_3}^{\sigma_1',\sigma_2',\sigma_3'}
\to
\widetilde{\rop}\,\mystrut_{\sigma_1,\sigma_2,\sigma_3}^{\sigma_1',\sigma_2',\sigma_3'}(\as_1,\as_2,\as_3)=\\[.3cm]
\ds
\qquad\qquad
\;=\;\F(\tfrac{\eta}{4}-\as_1,\sigma_1+\sigma_1')\;
\F(\tfrac{\eta}{4}-\as_2,-\sigma_2-\sigma_2')\;\F(\tfrac{\eta}{4}-\as_3,
\sigma_3+\sigma_3')
\;\rop\mystrut_{\sigma_1,\sigma_2,\sigma_3}^{\sigma_1',\sigma_2',\sigma_3'}\;,
\end{array}
\label{R-tilda}
\ee
which also lead to a continuous (three-parameter) family of the
commuting transfer matrices. Indeed, the tetrahedron equation
\be
\begin{array}{l}
\widetilde{\rop}_{123}(\as_1,\as_2,\as_3)\, 
\widetilde{\rop}_{145}(\as'_1,\as'_2,\as'_3)\, 
\widetilde{\rop}_{246}(\as''_1,\as''_2,\as_3'-\as_3)\, 
\widetilde{\rop}_{356}(\as''_1-\as_1',
\as''_2-\as_1,\as'_2-\as_2)
\\[.7cm]
\quad=\widetilde{\rop}_{356}(\as''_1-\as_1',
\as''_2-\as_1,\as'_2-\as_2)\, 
\widetilde{\rop}_{246}(\as''_1,\as''_2,\as_3'-\as_3)\, 
\widetilde{\rop}_{145}(\as'_1,\as'_2,\as'_3)\, 
\widetilde{\rop}_{123}(\as_1,\as_2,\as_3),
\end{array}
\ee
\noindent
holds 
for arbitrary values of $\as_1,\as_2,\as_3$ and $\as_1',\as_2',\as_3'$.
Finally, define the field-dependent partition function
\be\label{Z-field}
Z^{\rm (field)}=
\sum_{\rm{spins}} \ \prod_{\rm{vertices}}\,
\widetilde{\rop}\,\mystrut_{\sigma_1,\sigma_2,\sigma_3}^{\sigma_1',
\sigma_2',\sigma_3'}(\as_1,\as_2,\as_3)\,.
\ee
Note, that particular cases of the solutions \eqref{W-im2} and
\eqref{qR1} with the Faddeev quantum dilogarithm were previously
obtained in \cite{Sergeev:1999jpa,BMS08a}.


\section{Examples of the quantum dilogarithms\label{examples}}
The notion of the quantum dilogarithm 
was introduced by Faddeev and
Kashaev \cite{Faddeev:1993rs} and further refined in \cite{Andersen:2014aoa,
Kashaev:2015,Garoufalidis:2023dhd}.   
In this section we review a few examples of quantum dilogarithms
in order to facilitate their application to the construction of integrable  
lattice models in three dimensions. 
\subsection{Faddeev dilogarithm\label{fad-sect}}
In this case Pontryagin group coincides with the real line
\be 
\S={\mathbb R}\,,\qquad \int_\S \rd x=\int_{\mathbb R}\rd x\,.
\ee
The Faddeev \cite{Faddeev:1994fw} quantum 
dilogarithm $\Phi_{\bb}(x)$ is a
meromorphic function in the complex plane $x\in{\mathbb C}$ 
defined by the product formula
\begin{equation}\label{fad-prod}
\Phi_{\bb}(x)\,=\,\frac{(-q\,e^{2\pi x{}\bb}\,;\ q^2)_\infty}
{( -\tilde q\,e^{2\pi x{}\bb^{-1}};\tilde  q^{\,2})_\infty},\qquad
q=\EXP^{\ii\pi \bb^2},\qquad \tilde{q}=\EXP^{-\ii\pi \bb^{-2}},\qquad
{\Im}m\, \bb^2>0\,,
\end{equation}
where $\bb$ is a free parameter and 
\begin{equation}\label{poch}
(x;q)_\infty\,=\,\prod_{k=0}^\infty (1-q^k x)\,,
\end{equation}
denotes the $q$-Pochhammer symbol. 
There are also integral representations, in particular,
the one given by Faddeev \cite{Faddeev:1994fw}, 
with the integral over the real line
${\mathbb R}$, 
\begin{equation}\label{fad-int}
\Phi_{\bb}(x)\,=\,\exp\left\{\frac{1}{4}\int_{\mathbb{R}+\ii 0} \frac{\ds \EXP^{-2\ii x z}}{\ds \sinh(\bb z)\sinh(\bb^{-1}z)} \frac{dz}{z}\right\}\,,\qquad
\end{equation}
which is convenient for the analytic continuation of $\Phi_{\bb}(x)$
to real values of $\bb$. 

The definition \eqref{fad-prod} implies that
$\Phi_{\bb}(x)$ is meromorphic function of $x$ with simple zeroes at 
$x= -(\eta_\bb + \ii m b +\ii n b^{-1})$
and simple poles at 
$x=\eta_\bb 
+ \ii m b + \ii n b^{-1}$, where $m,n\in\mathbb{Z}_{\geq 0}$ and 
\be\label{eta-def}
\eta_\bb=\ri(\bb+\bb^{-1})/2\,.
\ee
It has the following asymptotics
\begin{equation}
\Phi_\bb(x)\;\simeq\;1,\qquad\Re e(x)\to
-\infty;\qquad\Phi_\bb(x)\;\simeq\;\EXP^{\ii\pi
x^2-\ii\pi(1-2\eta_\bb^2)/6}\qquad \Re e(x)\to
+\infty\;,
\end{equation}
where $\Im m(x)$ is kept finite.

\bigskip
In line with the general notation of Sect.\ref{s22}, 
let us now set the quantum dilogarithm and
the matrix elements of the Fourier transform operator and 
the Gaussian exponent, 
defined in \eqref{kernel} and \eqref{gauss-def},
\begin{subequations}\label{fad-case} 
\begin{equation}\label{fad-case1}
\varphi(x)=\Phi_\bb(x)\,,\qquad \F(x,y)=\re^{2\pi\ri x y}\,,
\qquad \G(x)=\re^{\ri\pi x^2}\,,\qquad x,y\in {\mathbb R}\,,
\ee
as well as the related constants
\be\label{phib0-def}
\varphi(0)^2={\Phi}_{\bb}(0)^2=\re^{-\ri\pi\eta_\bb^2/3-\ri\pi/6}\,,
\qquad\qquad \go=\int_{\mathbb R} \re^{\ri \pi x^2}\rd x=
\re^{\ri\pi/4}\,,\qquad\qquad \eta=\eta_\bb\,.
\ee
Note, that these constants satisfy the relation \eqref{parrel}. 
\end{subequations}
Now, one can check, 
that the Faddeev dilogarithm satisfies all the general relations 
\eqref{inv-gen}-\eqref{fterm3} for the Fourier self-dual quantum dilogarithm 
with $\S={\mathbb R}$.  An analytical proof of the quantum five-term
identity 
\eqref{fterm3} could be found in \cite{FKV:2001}.
Note, that in the quasiclassical limit $\bb\to 0$ this identity
reduces to the classical five-term identity for
the Rogers dilogarithm, see, e.g., \cite{Kashaev2011}.

\subsection{Andersen--Kashaev dilogarithm}
In this case the Pontryagin group $\S={\mathbb R}\times {\mathbb Z}_N$
has the structure of a direct product 
\begin{equation}\label{Kashaev1}
\S=\mathbb{R}\times \mathbb{Z}_N\,,\qquad
x=(\xi,n)\in \S\;,
\qquad \int_{\S}\rd x= \frac{1}{\sqrt{N}}\, 
\sum_{n\in\mathbb{Z}_N} \int_{\mathbb{R}} d\xi\;,
\end{equation}
where $\mathbb{Z}_N$ is the set of integers modulo $N$ ($N\ge2$).
The elements $x\in\S$ have 
two components $x=(\xi,n)$ where $\xi\in {\mathbb R}$ and $n\in
{\mathbb Z}_N$.
Similarly to \eqref{fad-prod} and \eqref{eta-def} introduce the notations
\be\label{eta-def2}
q=\EXP^{\ii\pi \bb^2}\,,\qquad \tilde{q}=\EXP^{-\ii\pi
  \bb^{-2}}\,,\qquad \eta_\bb=\ri(\bb+\bb^{-1})/2\,,
\ee
where $\bb$ is a free parameter. 
A solution of the %
defining relation
\eqref{inv-gen}-\eqref{fterm3} for this case was obtained by Andersen
and Kashaev \cite{Andersen:2014aoa}
\begin{equation}
\varphi(x)\;=\;\prod_{j=0}^{N-1} \Phi_\bb\left(\frac{\xi}{\sqrt{N}}+
(1-\frac{1}{N})\eta_\bb-\ii\bb^{-1}\frac{j}{N}-\ii\bb
\left\{\frac{j+n}{N}\right\}\right)\;,\qquad x=(\xi,n)\,,  
\end{equation}
where the symbol
$\{\ \}$ denotes the positive fractional part, while 
$\Phi_\bb(\xi)$ is defined by \eqref{fad-prod}. This formula implies the
product representation,
\begin{equation}\label{prodrep}
\varphi(x)=\frac{(\omega^n\,\re^{2\pi\bb(\xi\sqrt{N}+\eta_\bb)/N};\omega
 \, {q}^{2/N})_\infty}{(\omega^{-n}\,\re^{2\pi\bb^{-1}(\xi\sqrt{N}-\eta_\bb)/N};\omega^{-1}
  {\tilde{q}}^{\,2/N})_\infty}\,,\quad x=(\xi,n),
\quad \omega=\re^{2\pi \ri/N}\,,\quad
{\Im}m\, \bb^2>0.
\end{equation}
The matrix elements of the operators $\Fop$ and $\G(\xop)$ read
\be
\F(x,x')=\re^{2\pi \ri \xi \xi'}\,\re^{-2\pi \ri n n'/N},
\quad \G(x)=\EXP^{\ii\pi\xi^2}\EXP^{-\ii\pi n
  (n+N)/N},\quad
x=(\xi,n)\,,\quad x'=(\xi',n')\,,
\ee
and the other related constants are given by 
\begin{equation}
\go=\EXP^{\ii\pi N/4}\,,\qquad 
\varphi(0)^2=\EXP^{-\ii\pi(N+2\eta_\bb^2/N)/6}\,,
\qquad \eta\;=\;(\frac{\eta_\bb}{\sqrt{N}},0)\,.
\end{equation}

\subsection{Woronowicz dilogarithm}
In this case the Pontryagin group $\S=\mathbb{T}\times\mathbb{Z}$ also
has the structure of direct product
\be\label{dim-def}
\S=\mathbb{T}\times\mathbb{Z}\,,\qquad
x=(\theta,m)\,,\qquad 
\qquad \int_{\S}\rd x=\frac{1}{2\pi}\,
\sum_{m\in\mathbb{Z}} \int_0^{2\pi} d\theta\,,
\ee
where $\mathbb{T}$ denotes the unit circle and $\mathbb{Z}$ denotes
the set of all integers. The elements $x\in\S$ have 
two components $x=(\theta,m)$, where $0\le\theta<2\pi$ and $m\in
{\mathbb Z}$.
A solution of \eqref{inv-gen}-\eqref{fterm3}
for this case was obtained by Woronowicz \cite{Woronowicz:1992} (see also 
\cite{Dimofte:2011py}), 
\be\label{dgg-def}
\ph(x)\;=\;\frac{(-q^{1-m}\EXP^{\ii\theta};q^2)_\infty}{(-q^{1-m}\EXP^{-\ii\theta};q^2)_\infty}\,,\qquad
q=-\re^{\ri \eta_0}\,,
\qquad |q|<1\,,
\end{equation}
where $\eta_0$ is a free parameter.
The matrix elements of the operators $\Fop$ and $\G(\xop)$, defined in \eqref{kernel} and \eqref{gauss-def}, are given by
\begin{equation}
\F(x,x')=\re^{\ri\theta m'+\ri\theta' m}\,,
\qquad \G(x)=\re^{\ri\theta m}\,,\qquad x=(\theta,m)\,,\qquad
x'=(\theta',m')\,.
\ee
and the relevant constants read
\be
\go=1\,,\qquad \varphi(0)^2=1\,,\qquad 
\eta=(\eta_0,0)\,.
\end{equation}
\section{Partition functions of 3D models\label{partition}}
In this section we report the results of calculations \cite{bkms2}
for the partition functions of the 3D models, associated with the
Faddeev dilogarithm. Therefore, in what follows we will always assume
the substitutions \eqref{fad-case} for the quantum dilogarithm
$\varphi(x)={\Phi}_{\bb}(x)$
 and all related quantities. Moreover, for definiteness, we
will consider the regime when the parameter $\bb$ is such that  
\be\label{regime}
\Im m\,\bb=0\,,\qquad\mbox{or}\qquad |\bb|=1\,,
\ee
with $\Re e\,\bb>0$.
In this case the quantum dilogarithm function 
$\varphi(x)={\Phi}_{\bb}(x)$ becomes unimodular for
real values of $x$, 
\be\label{unimod}
|\varphi(x)|=1\,,\qquad \forall x\in{\mathbb R}\,.
\ee

\subsection{The 3D vertex model\label{p-vertex}}
The partition function of the model on the $M\times M\times M$ cubic lattice
is defined by \eqref{Zvertex}. The vertex weights
are given by \eqref{R-matrix2} and \eqref{ro-1} (with the substitutions
\eqref{fad-case})  
with the edge spins taking
real values $x_1,x_1'\ldots \in {\mathbb R}$.
As before, we assume periodic boundary conditions in all three
directions. 
The vertex weights \eqref{R-matrix2} depend on three spectral parameters
$\lambda_1,\lambda_2,\lambda_3\in {\mathbb R}$. 
Let %
\be\label{z-vert-inf}
 z^{\rm (vert)}_\infty(\lambda_1,\lambda_2,\lambda_3)
=\lim_{M\to \infty}\,
\big(Z^{\rm (vert)}\big)^{1/M^3}
\ee
denote the
partition function per site 
in the 
limit of an infinitely large lattice. 

To concisely present the exact expression for the partition function
\eqref{z-vert-inf} one needs to use the ``angle parameterization'' of
the $\lambda$'s in
terms of elements of a spherical triangle.  Let
$\theta_1,\theta_2,\theta_2$ denote the angles of a spherical triangle
and $a_1, a_2, a_3$ denote the lengths of the sides (or the linear angles)
of this triangle opposite to 
$\theta_1,\theta_2,\theta_2$, respectively. 
The two sets of angles are connected by the 
spherical cosine theorem,
\begin{equation}\label{cos-t}
\cos a_i\;=\;
\frac{\cos\theta_i+\cos\theta_j\cos\theta_k}
     {\sin\theta_j\sin\theta_k}\;,\qquad  (i,j,k)={\rm perm}(1,2,3)\,.
\end{equation}
Define related variables
\begin{equation}\label{beta}
\beta_0\;=\;\pi-\frac{a_1+a_2+a_3}{2}\;,\quad \beta_j\;=\;\pi-\beta_0-a_j\;,
\qquad j=1,2,3\,.
\end{equation}
and consider the case when $\theta_1,\theta_2,\theta_2$ and
$\beta_0,\beta_1,\beta_2,\beta_3$ are all real and between $0$ and $\pi$.
We can now introduce the ``angle parameterization'' 
\begin{equation}\label{lam-th}
\EXP^{\ii\frac{\pi}{\eta}\lambda_1}\;=\;\tan^2\frac{\theta_1}{2}\;,\quad 
\EXP^{\ii\frac{\pi}{\eta}\lambda_2}\;=\;\cot^2\frac{\theta_2}{2}\;,\quad
\EXP^{\ii\frac{\pi}{\eta}\lambda_3}\;=\;\tan^2\frac{\theta_3}{2}\;.
\end{equation}
or, equivalently, as 
\begin{equation}\label{tan}
\EXP^{\ii\frac{\pi}{\eta}\lambda_1}\;=\;\frac{\sin\beta_2\sin\beta_3}{\sin\beta_0\sin\beta_1}\;,\quad 
\EXP^{\ii\frac{\pi}{\eta}\lambda_2}\;=\;\frac{\sin\beta_0\sin\beta_2}{\sin\beta_1\sin\beta_3}\;,\quad 
\EXP^{\ii\frac{\pi}{\eta}\lambda_3}\;=\;\frac{\sin\beta_1\sin\beta_2}{\sin\beta_0\sin\beta_3}\;.
\end{equation}
Note, that for the regime \eqref{regime}, the quantity $\eta=\eta_\bb$
is positive imaginary, and for real $\theta$'s the corresponding
parameters $\lambda$'s are real.
With this parameterization, 
the partition function per site \eqref{z-vert-inf} can be
written as \cite{bkms2}, 
\begin{equation}\label{pf-2}
\log z_\infty^{\rm vert}\;=\; \frac{4\eta^2}{\pi}\sum_{j=0}^3\Big( 
\Lambda(\beta_j)+(\beta_j-\frac{\pi}{4})\log(2 \sin \beta_j)\Big)\;,
\qquad \eta=\ri(\bb+\bb^{-1})/2
 \end{equation}
where 
\be\label{Lob}
\Lambda(\beta)\;=\;-\int_0^\beta\log (2\sin x) \,\rd x\;,
\ee
is the Lobachevsky function. It is worth noting that the modular
parameter $\bb$ enters the quantum dilogarithm \eqref{fad-int} and,
hence, the Boltzmann weights \eqref{R-matrix2}, 
in a rather complicated way. Contrary to
this, the partition function \eqref{pf-2} depends on $\bb$  
in a very simple way: this parameter only 
influences the coefficient in front of
the free energy.

Finally note, that at the symmetric point
\be
\lambda_1=\lambda_2=\lambda_3=0\quad\Rightarrow\quad
\theta_1=\theta_2=\theta_3=a_1=a_2=a_3=\frac{\pi}{2}\,,
\ee
the partition function takes the form
 \begin{equation}
z_\infty^{\rm vert}\;=\;\exp\biggl( \frac{8\eta^2}{\pi}G\biggl)\;,
 \end{equation}
where $G = 0.915965\ldots$ is the Catalan constant. Here we have used
the value $\Lambda(\tfrac{\pi}{4})=\tfrac{1}{2}G$. 

\subsection{The interaction-round-a-cube (IRC) model\label{p-irc}}
Consider now the IRC model on the $M\times M\times M$ cubic lattice
with the Boltzmann weights $W(\ldots)$
given by \eqref{W-im2}, \eqref{ro-w} (with the substitutions
\eqref{fad-case}) and the partition function 
defined by \eqref{Z-cube}. We assume the periodic boundary conditions for the
site spins. Similarly to \eqref{z-vert-inf} define the partition
function per site in the limit of the infinite lattice
\be
\label{z-irc-inf}
z^{\rm (IRC)}_\infty(\lambda_1,\lambda_2,\lambda_3)=
\lim_{M\to \infty}\,\big(Z^{\rm (IRC)}\big)^{1/M^3}\,.
\ee
As shown in \cite{bkms2} this partition function exactly coincides 
with that of the vertex model 
\be
z^{\rm (IRC)}_\infty(\lambda_1,\lambda_2,\lambda_3)=
z^{\rm vert}_\infty(\lambda_1,\lambda_2,\lambda_3)\,.
\ee
presented in \eqref{pf-2} of the previous
subsection. 

It is interesting to compare the above result with the partition
function of the Zamolodchikov--Bazhanov--Baxter model (which is the
generalization \cite{Bazhanov:1992jq,Bazhanov:1993j}
of the original Zamolodchikov model
\cite{Zamolodchikov:1980rus,Zamolodchikov:1981kf} for $N\ge2$ discrete
spin states). Denoting this partition function per site as $\kappa_N$, one
can write 
it in the form\footnote{See eqs.(3.32)-(3.34) in \cite{Bazhanov:1993j}}
\be\label{pf-zbb}
\log\kappa_N =\frac{2 (N-1)}{N\pi}\sum_{j=0}^3
\Big(\Lambda(\beta_j)+\frac{\beta_j}{2}\log(2
\sin\beta_j)-\frac{\pi}{2}
\log\big(\sqrt{2}\cos({\beta_j}/{2})\big)\Big)\,, 
\ee
which is, evidently, very similar to \eqref{pf-2} (the parameters
$\beta_j$ are the same). Note also, that the number of spin
states $N$ only enters \eqref{pf-zbb} through an overall factor in
front of the free energy.  

\subsection{The vertex model with external fields}
Consider now the vertex model with external fields, whose  
partition function $Z^{\rm field}$ 
is defined by \eqref{Z-field} with the vertex weights
given by \eqref{qR1} and \eqref{R-tilda}. As before, we consider the model
on the cubic lattice of the size 
$M\times M\times M$ with the periodic boundary
conditions for the edge spins in all three directions. 

Using the relations \eqref{sigma} and
\eqref{R-WS} one can express the partition function $Z^{\rm field}$
as an integral of a spectral parameter inhomogeneous version of the 
partition function of the IRC model \eqref{Z-cube} 
over a set of $(6M-3)$ spectral
parameters. The connection is exact for a finite lattice. The details
are presented in \cite{bkms2}. In the infinite lattice limit the
integral can be taken by the saddle point method. The result 
is 
\be\label{pf-field}
z^{\rm field}_\infty\;=\;\lim_{M\to\infty}\;
\big(Z^{\rm (field)}\big)^{1/M^3}\;=\;
\exp\biggl( \frac{4\eta^2}{\pi}\sum_{j=0}^3 \Lambda(\beta_j)\biggl)\;,
\ee
where $\Lambda(\beta)$ is given by \eqref{Lob} and 
$\beta$'s are defined in \eqref{beta} with the linear angles  
\be
a_i=\frac{2\pi \phi_i}{\eta}\,,\qquad i=1,2,3.
\ee
determined by the external fields $\phi_1,\phi_2,\phi_3$ 
entering the definitions
\eqref{Z-field} and \eqref{R-tilda}.    

\section{Conclusion}
In this paper we introduced new integrable 3D lattice models. Their
construction is based on the algebraic properties of the quantum
dilogarithms.
In particular, we presented general formulae \eqref{elem-gen},
\eqref{W-im2} expressing a substantial class of solutions of the
Zamolodchikov tetrahedron equation in terms of quantum dilogarithms,
both for the ``interaction-round-a-cube'' (IRC) and the vertex-type models.
A detailed proof of the Zamolodchikov tetrahedron equation for these
solutions, as well as an explanation of the connection between the
3D vertex and IRC models is contained in our detailed paper \cite{bkms2}. 
The proof presented there 
is rather general: it is only based on the
basic properties of the quantum dilogarithms, given by \eqref{inv-gen}
and \eqref{five1}.
Three different examples of the quantum
dilogarithm possessing these properties were reviewed 
above in Sect.\ref{examples} of this paper.

The local spin variables 
in the corresponding lattice models run over some specific (model
dependent) sets of values, identified with Pontryagin
self-dual locally compact Abelian (LCA) groups.
In the simplest case, associated with the Faddeev modular quantum
dilogarithm, the spins take arbitrary real values. The correponding
vertex and IRC models contain a free modular parameter $\bb$\  ($\Im m
\bb^2\ge0$) and three real spectral parameters
$\lambda_1,\lambda_2,\lambda_3$, which are conveniently parameterized 
\eqref{tan} through the linear accesses
$\beta_0,\beta_1,\beta_2,\beta_3$ of the spherical triangle, defined
in 
\eqref{beta}. The exact results for the partition functions
per site for these models are presented in
\eqref{pf-2} and \eqref{pf-field}, whereas details of calculations are
contained in our forthcoming paper \cite{bkms2}. 
Naturally, the appearance of the
Lobachevsky function $\Lambda(\beta)$ (see \eqref{Lob}) in these
results suggests connections to problems of geometry. Indeed, the
quasi-classical (or the low-temperature) 
limit $\bb\to 0$ in the above lattice models is related to the
problem of constructing of {\em circular quadraliteral
  lattices} in $3D$ space --- lattices whose faces are planar
quadrilaterals inscribable into a circle \cite{Bazhanov:2008rd,Bobenko:1999,
DoliwaSantini,KS98}. 
We intend to discuss
this connection in our future article \cite{bkms2}.

\section*{Acknowledgements} 
One of us (VVB) thanks S.L.Lukyanov and A.B.Zamolodchikov for
important comments.

\bibliography{total33}

\newcommand\oneletter[1]{#1}\def\cprime{$'$}
\begin{thebibliography}{10}

\bibitem{Zamolodchikov:1980rus}
Zamolodchikov, A.~B.
\newblock Tetrahedra equations and integrable systems in three-dimensional
  space.
\newblock Soviet Phys. JETP {\bf 52} (1980) 325--336.

\bibitem{Zamolodchikov:1981kf}
Zamolodchikov, A.~B.
\newblock Tetrahedron equations and the relativistic {S} matrix of straight
  strings in (2+1)-dimensions.
\newblock Commun. Math. Phys. {\bf 79} (1981) 489--505.

\bibitem{Bazhanov:1981zm}
Bazhanov, V.~V. and Stroganov, Y.~G.
\newblock On commutativity conditions for transfer matrices on multidimensional
  lattice.
\newblock Theor. Math. Phys. {\bf 52} (1982) 685--691.

\bibitem{Bax82}
Baxter, R.~J.
\newblock {\em Exactly {S}olved {M}odels in {S}tatistical {M}echanics}.
\newblock Academic, London, 1982.

\bibitem{Baxter:1983qc}
Baxter, R.~J.
\newblock On {Z}amolodchikov's solution of the tetrahedron equations.
\newblock Commun. Math. Phys. {\bf 88} (1983) 185--205.

\bibitem{Bazhanov:1981eg}
Bazhanov, V.~V. and Stroganov, Y.~G.
\newblock {Free Fermions on Three-dimensional Lattice and Tetrahedron
  Equations}.
\newblock Nucl. Phys. B {\bf 230} (1984) 435--454.

\bibitem{Baxter:1986phd}
Baxter, R.~J.
\newblock The {Y}ang-{B}axter {E}quations and the {Z}amolodchikov {M}odel.
\newblock Physica {\bf 18D} (1986) 321--247.

\bibitem{Maillet:1989gg}
Maillet, J.~M. and Nijhoff, F.
\newblock Integrability for multidimensional lattice models.
\newblock Phys. Lett. {\bf B224} (1989) 389.

\bibitem{Bazhanov:1992jq}
Bazhanov, V.~V. and Baxter, R.~J.
\newblock New solvable lattice models in three-dimensions.
\newblock J. Statist. Phys. {\bf 69} (1992) 453--585.

\bibitem{Bazhanov:1993j}
Bazhanov, V.~V. and Baxter, R.~J.
\newblock Star triangle relation for a three-dimensional model.
\newblock J. Statist. Phys. {\bf 71} (1993) 839--864.

\bibitem{Kashaev:1993ijmp}
Kashaev, R.~M., Mangazeev, V.~V., and Stroganov, Y.~G.
\newblock Spatial symmetry, local integrability and tetrahedron equations in
  the {B}axter-{B}azhanov model.
\newblock Int. J. Mod. Phys. A {\bf 8} (1993) 587.

\bibitem{Korepanov:1993jsp}
Korepanov, I.~G.
\newblock Tetrahedral {Z}amolodchikov algebras corresponding to {B}axter's
  ${L}$-operators.
\newblock Comm. Math. Phys. {\bf 154} (1993) 85--97.

\bibitem{Bazhanov:1993pa}
Bazhanov, V.
\newblock Inversion and symmetry relations for a three-dimensional solvable
  model.
\newblock {I}nt. J. Mod. Phys. B {\bf 07} (1993) 3501--3515.

\bibitem{Hietarinta:1994pt}
Hietarinta, J.
\newblock Labeling schemes for tetrahedron equations and dualities between
  them.
\newblock J. Phys. {\bf A27} (1994) 5727--5748.

\bibitem{Kapranov94}
Kapranov, M.~M. and Voevodsky, V.~A.
\newblock 2-categories and Zamolodchikov tetrahedra equations.
\newblock Proceedings of Symposia in Pure Mathematics {\bf 56} (1994) 177--259.

\bibitem{Sergeev:1995rt}
Sergeev, S.~M., Mangazeev, V.~V., and Stroganov, Y.~G.
\newblock The vertex formulation of the {B}azhanov-{B}axter model.
\newblock J. Stat. Phys. {\bf 82} (1996) 31--50.

\bibitem{Kashaev:1996on}
Kashaev, R.~M.
\newblock On discrete three-dimensional equations associated with the local
  {Y}ang-{B}axter relation.
\newblock Lett. Math. Phys. {\bf 38} (1996) 389--397.

\bibitem{Baxter:1997tw}
Baxter, R.~J. and Bazhanov, V.~V.
\newblock Two-layer {Z}amolodchikov model.
\newblock In {\em XIIth International Congress of Mathematical Physics (ICMP
  ’97) (Brisbane)}, pages 15--23. Int. Press, Cambridge, MA, 1999.

\bibitem{Sergeev:1998so}
Sergeev, S.~M.
\newblock Solutions of the functional tetrahedron equation connected with the
  local {Y}ang-{B}axter equation for the ferro-electric condition.
\newblock Lett. Math. Phys. {\bf 45} (1998) 113--119.

\bibitem{MR1637789}
Kashaev, R.~M. and Sergeev, S.~M.
\newblock On pentagon, ten-term, and tetrahedron relations.
\newblock Comm. Math. Phys. {\bf 195} (1998) 309--319.

\bibitem{Kashaev:2000fr}
Kashaev, R.~M. and Volkov, A.~Y.
\newblock From the tetrahedron equation to universal {$R$}-matrices.
\newblock In {\em L. {D}. {F}addeev's {S}eminar on {M}athematical {P}hysics},
  volume 201 of {\em Amer. Math. Soc. Transl. Ser. 2}, pages 79--89. Amer.
  Math. Soc., Providence, RI, 2000.

\bibitem{Bazhanov:2005as}
Bazhanov, V.~V. and Sergeev, S.~M.
\newblock {Zamolodchikov's tetrahedron equation and hidden structure of quantum
  groups}.
\newblock J. Phys. {\bf A39} (2006) 3295--3310.

\bibitem{BMS08a}
Bazhanov, V.~V., Mangazeev, V.~V., and Sergeev, S.~M.
\newblock Quantum geometry of 3-dimensional lattices.
\newblock J. Stat. Mech  (2008) P07004.

\bibitem{Sergeev:2008zu}
Sergeev, S.~M.
\newblock {Super-tetrahedra and super-algebras}.
\newblock J. Math. Phys. {\bf 50} (2009) 083519.

\bibitem{Inoue:2023rer}
Inoue, R., Kuniba, A., and Terashima, Y.
\newblock {Tetrahedron equation and quantum cluster algebras}.
\newblock J. Phys. A {\bf 57} (2024) 085202.

\bibitem{KOS:2014}
Kuniba, A., Okado, M., and Sergeev, S.
\newblock Tetrahedron {E}quation and {Q}uantum {R} Matrices for modular double
  of ${U}_q({D}^{(2)}_{n+1})$, ${U}_q({A}^{(2)}_{2n})$ and
  ${U}_q({C}^{(1)}_{n})$.
\newblock arXiv:1409.1986, 2014.

\bibitem{Sergeev:2004pn}
Sergeev, S.~M.
\newblock {Q}uantum integrable models in discrete 2+1 dimensional space-time:
  auxiliary linear problem on a lattice, zero curvature representation,
  isospectral deformation of the {Z}amolodchikov-{B}azhanov-{B}axter model.
\newblock Particles and Nuclei {\bf 35} (2004) 1051--1111.

\bibitem{Faddeev:1993rs}
Faddeev, L.~D. and Kashaev, R.~M.
\newblock Quantum Dilogarithm.
\newblock Mod. Phys. Lett. {\bf A9} (1994) 427--434.

\bibitem{bkms2}
Bazhanov, V.~V., Kashaev, R.~M., Mangazeev, V.~V., and Sergeev, S.~M.
\newblock Quantum dilogarithms and new integrable lattice models in three
  dimensions. {II}. {A}lgebraic structures and partition function.
\newblock to be published, 2026.

\bibitem{Faddeev:1994fw}
Faddeev, L.~D.
\newblock {Current - like variables in massive and massless integrable models}.
\newblock In {\em {International School of Physics 'Enrico Fermi': 127th
  Course: Quantum Groups and Their Physical Applications}}, pages 117--136,
  1994.

\bibitem{Andersen:2014aoa}
Andersen, J.~E. and Kashaev, R.
\newblock {Complex Quantum Chern-Simons}.
\newblock arXiv:1409.1208, 2014.

\bibitem{Woronowicz:1992}
Woronowicz, S.~L.
\newblock Operator equalities related to the quantum {$E(2)$} group.
\newblock Comm. Math. Phys. {\bf 144} (1992) 417--428.

\bibitem{MR0005741}
Weil, A.
\newblock {\em L'int\'egration dans les groupes topologiques et ses
  applications}.
\newblock Actual. Sci. Ind., no. 869. Hermann et Cie., Paris, 1940.
\newblock [This book has been republished by the author at Princeton, N. J.,
  1941.].

\bibitem{Kashaev:2015}
Kashaev, R.
\newblock The {Y}ang-{B}axter relation and gauge invariance.
\newblock J. Phys. A {\bf 49} (2016) 164001.

\bibitem{Sergeev:2009jgp}
Sergeev, S.~M.
\newblock Geometry of quadrilateral nets: Second Hamiltonian form.
\newblock Journal of Geometry and Physics {\bf 59} (2009) 1150--1154.

\bibitem{Sergeev:1999jpa}
Sergeev, S.~M.
\newblock {Q}uantum {$2+1$} evolution model.
\newblock J. Phys. A: Math. Gen. {\bf 32} (1999) 5693--5714.

\bibitem{Bax73b}
Baxter, R.~J.
\newblock Eight-vertex model in lattice statistics and one-dimensional
  anisotropic {H}eisenberg chain. {II}. {E}quivalence to a generalized
  {I}ce-type lattice model.
\newblock Ann. Phys. {\bf 76} (1973) 25--47.

\bibitem{Garoufalidis:2023dhd}
Garoufalidis, S. and Kashaev, R.
\newblock {Quantum dilogarithms over local fields and invariants of
  3-manifolds}.
\newblock arXiv:2306.01331, 2023.

\bibitem{FKV:2001}
Faddeev, L.~D., Kashaev, R.~M., and Volkov, A.~Y.
\newblock Strongly coupled quantum discrete Liouville theory. I. Algebraic
  approach and duality.
\newblock Commun. Math. Phys. {\bf 219} (2001) 199--219.

\bibitem{Kashaev2011}
Kashaev, R.~M. and Nakanishi, T.
\newblock Classical and Quantum Dilogarithm Identities.
\newblock SIGMA {\bf 7} (2011) 102.

\bibitem{Dimofte:2011py}
Dimofte, T., Gaiotto, D., and Gukov, S.
\newblock {3-Manifolds and 3d Indices}.
\newblock Adv. Theor. Math. Phys. {\bf 17} (2013) 975--1076.

\bibitem{Bazhanov:2008rd}
Bazhanov, V.~V., Mangazeev, V.~V., and Sergeev, S.~M.
\newblock {Quantum geometry of 3-dimensional lattices}.
\newblock J. Stat. Mech. {\bf 0807} (2008) P07004.

\bibitem{Bobenko:1999}
Bobenko, A.~I.
\newblock Discrete conformal maps and surfaces.
\newblock London Math. Soc. Lecture Note Ser. {\bf 255} (1999) 97--108.

\bibitem{DoliwaSantini}
Doliwa, A. and Santini, P.~M.
\newblock Multidimensional quadrilateral lattices are integrable.
\newblock Phys. Lett. A {\bf 233} (1997) 365--372.

\bibitem{KS98}
Konopelchenko, B.~G. and Schief, W.~K.
\newblock Three-dimensional integrable lattices in {E}uclidean spaces:
  conjugacy and orthogonality.
\newblock R. Soc. Lond. Proc. Ser. A Math. Phys. Eng. Sci. {\bf 454} (1998)
  3075--3104.

\end{thebibliography}
\bibliographystyle{vvb-bibstyle}
\end{document}